\documentclass[%
 reprint,
superscriptaddress,
amsmath,amssymb,
aps,
]{revtex4-2}

\usepackage{graphicx}
\usepackage{dcolumn}
\usepackage{bm}

\usepackage{enumitem}  
\DeclareMathAlphabet\mathbfcal{OMS}{cmsy}{b}{n}
\newcommand{\of}[1]{\!\left(#1\right)}
\newcommand{\sqof}[1]{\left[#1\right]}
\newcommand{\cuof}[1]{\left\{#1\right\}}

\newcommand{\itt}[1]{ { \it{#1}}}

\def\kk{\boldsymbol{k}}
\def\vv{\boldsymbol{v}}
\def\rr{\boldsymbol{r}}
\def\BB{\boldsymbol{B}}

\def\EE{\boldsymbol{E}}
\def\JJ{\boldsymbol{J}}
\def\lll{\boldsymbol{l}}
\def\Bnabla{\boldsymbol{\nabla}}

\usepackage{float}
\usepackage{hyperref} 
\usepackage{cleveref} 

\hypersetup{colorlinks = true, 
	    linkcolor = blue, 
	    urlcolor = blue,
            citecolor = blue} 

\begin{document}

\preprint{APS/123-QED}

\title{First principles methodology for studying  magnetotransport in narrow-gap semiconductors: an application to Zirconium Pentatelluride $\mathrm{ZrTe_5}$}
        \author{Hanqi Pi}
	\affiliation{Beijing National Laboratory for Condensed Matter Physics and Institute of physics, Chinese academy of sciences, Beijing 100190, China}
	\affiliation{University of Chinese academy of sciences, Beijing 100049, China}
         \author{Shengnan Zhang}
         \email{shengnan.zhang@iphy.ac.cn} 
	\affiliation{Beijing National Laboratory for Condensed Matter Physics and Institute of physics, Chinese academy of sciences, Beijing 100190, China}
          \author{Yang Xu}
	\affiliation{Beijing National Laboratory for Condensed Matter Physics and Institute of physics, Chinese academy of sciences, Beijing 100190, China}
 	\affiliation{University of Chinese academy of sciences, Beijing 100049, China}
        \author{Zhong Fang} 
	\affiliation{Beijing National Laboratory for Condensed Matter Physics and Institute of physics, Chinese academy of sciences, Beijing 100190, China}
	\affiliation{University of Chinese academy of sciences, Beijing 100049, China}
	\author{Hongming Weng}
	\email{hmweng@iphy.ac.cn} 
	\affiliation{Beijing National Laboratory for Condensed Matter Physics and Institute of physics, Chinese academy of sciences, Beijing 100190, China}
	\affiliation{University of Chinese academy of sciences, Beijing 100049, China}
	\affiliation{Songshan Lake Materials Laboratory, Dongguan, Guangdong 523808, China}
	\author{Quansheng Wu}
	\email{quansheng.wu@iphy.ac.cn} 
	\affiliation{Beijing National Laboratory for Condensed Matter Physics and Institute of physics, Chinese academy of sciences, Beijing 100190, China}
	\affiliation{University of Chinese academy of sciences, Beijing 100049, China}

\date{\today}

\begin{abstract}
The origin of anomalous resistivity peak and accompanied sign reversal of Hall resistivity of ZrTe$_5$ has been under debate for a long time. Although various theoretical models have been proposed to account for these intriguing transport properties, a systematic study from first principles view is still lacking. In this work, we present a first principles calculation combined with Boltzmann transport theory to investigate the transport properties in narrow-gap semiconductors at different temperatures and doping densities within the relaxation time approximation. Regarding the sensitive temperature-dependent chemical potential and relaxation time of semiconductors, we take proper approximation to simulate these two variables, and then comprehensively study the transport properties of ZrTe$_5$ both in the absence and presence of an applied magnetic field. Without introducing topological phases and correlation interactions,  we qualitatively reproduced crucial features observed in experiments, including zero-field resistivity anomaly, nonlinear Hall resistivity with sign reversal, and non-saturating magnetoresistance at high temperatures. Our calculation allows a systematic interpretation of the observed properties in terms of multi-carrier and Fermi surface geometry. Our method can be extended to other narrow-gap semiconductors and further pave the way to explore interesting and novel transport properties of this field.

\end{abstract}

\maketitle


\section{Introduction}

The galvanomagnetic properties of solids, describing transport behaviors of charge carriers driven by electric and magnetic fields, have been intensively studied for their potential applications in both fundamental and industrial research. In the realm of fundamental scientific research, the galvanomagnetic response is a powerful tool to investigate the electronic structure\cite{tang2019three,zhu2015quantum,terashima2018fermi,alexandradinata2018revealing,chen2021extremely,luo2015hall,hou2015high}, topological properties\cite{son2013chiral,xiong2015evidence,hirschberger2016chiral,huang2015observation} and scattering mechanism\cite{orlita2011carrier,leuliet2006electron} of materials. In industrial applications, materials with strong magnetic responses are promising candidates for magnetometer\cite{heremans1993narrow,reig2009magnetic,granell2019highly,henriksen2010planar} and hard drives\cite{daughton1999gmr}.  However, the galvanomagnetic phenomena influenced by both intrinsic and extrinsic effects have distinct origins in various systems. This complexity makes it challenging to interpret certain magneto-transport behaviors despite extensive study. For instance, the extreme large magnetoresistance (XMR) can be attributed to nontrivial band topology\cite{he2014quantum,liang2015ultrahigh}, charge carrier compensation\cite{ali2014large,thirupathaiah2017mote} and open orbitals of charge carriers\cite{takatsu2013extremely,zhang2019magnetoresistance}. The planar Hall effect can arise from the strong spin-orbital coupling in magnetic material\cite{nazmul2008planar,tang2003giant,yin2019planar}, anisotropy of the Fermi surfaces\cite{yang2020large,liu2019nontopological,yang2019current,meng2019planar,liang2019origin} and the chiral anomaly in topological semimetals\cite{nandy2017chiral,burkov2017giant,kumar2018planar,li2018giant}. Furthermore, the nonlinearity of Hall resistivity might indicate the presence of multi-carrier behavior or suggest the occurrence of the anomalous Hall effect.

$\mathrm{ZrTe_5}$ has been extensively studied since the 1970s due to its remarkable thermoelectric properties and its resistivity anomaly accompanying sign reversal of Hall resistivity and Seebeck coefficients\cite{shahi2018bipolar,gourgout2022magnetic,wieting1980giant,okada1980giant,jones1982thermoelectric,mcilroy2004observation}. Plenty of mechanisms have been proposed to explain this anomalous peak, such as the Lifshitz transition\cite{manzoni2015ultrafast}, topological phase transition\cite{xu2018temperature}, formation of Dirac polarons\cite{fu2020dirac} and thermally excited charge carriers\cite{wang2021thermodynamically}. In addition, the nonlinear Hall resistivity has been regarded as the anomalous Hall effect (AHE) originating from the Berry curvature\cite{lozano2022anomalous,liang2018anomalous,liu2021induced,choi2020zeeman,wang2023theory}. Many other novel properties, such as the topological edge states\cite{manzoni2016evidence,wu2016evidence,zhang2017electronic,weng2014transition}, chiral magnetic effect\cite{liang2018anomalous,li2016chiral}, and quantum Hall effect\cite{tang2019three}, are believed to be connected with the nonzero Berry curvature in ZrTe$_5$. However, some experiment results cannot be explained by the Berry curvature mechanism. For example, the $\mathrm{Land\acute{e}}$ $g$-factor of ZrTe$_5$, ranging between 21 and 26\cite{liu2016zeeman,chen2015magnetoinfrared,sun2020large}, is insufficient to induce Weyl nodes through Zeeman splitting\cite{choi2020zeeman,liang2018anomalous} and hence generate the nonzero Berry curvature. Meanwhile, the multi-carrier model has proved to be efficient in some recent experiments and theoretical works\cite{shahi2018bipolar,liu2016zeeman,liu2023gate}.

Various models have been proposed to explain the unusual transport properties of ZrTe$_5$, yet there's a lack of comprehensive first principles studies that simulate without assuming any parameters. Moreover, ZrTe$_5$ is a typical narrow-gap semiconductor, of which the transport behaviors sensitively depend on the temperature and chemical potential. Therefore we employ first principles calculations combined with the Boltzmann transport theory to systematically investigate the galvanomagnetic property of ZrTe$_5$ at different doping densities and temperatures. We want to stress that the method for ZrTe$_5$ is more complicated than our previous one for metals/semimetals\cite{zhang2019magnetoresistance}. On the one hand, the chemical potential and charge carrier density of ZrTe$_5$ varies with the temperature sensitively. On the other hand, the relaxation time can not be obtained simply and directly as metals/semimetals. Therefore it is important to treat these two variables with proper approximations to enhance the alignment between the simulation and experiment, which is the innovation and necessity of this work. To achieve this, we first calculate the temperature-dependent chemical potential by fixing the total electron density, then obtain the product of temperature-dependent resistivity and relaxation time for different doping densities through an interpolation scheme. Finally, we fit the relaxation time using Bloch-Gr\"uneisen model.

Our method, without incorporating energy band topology effects or correlation interactions, successfully reproduces the essential experimental characteristics, including the anomalous peak of the temperature-dependent resistivity, the nonlinearity and sign reversal of Hall resistivity, and the non-saturating magnetoresistance (MR) at high temperatures\cite{liu2023gate,shahi2018bipolar,tritt1999large,zhou2019anisotropic,lozano2022anomalous}. More generally and importantly, our methodology can be extended to study other narrow-gap semiconductors. We believe our work provides invaluable insights into the unique behavior exhibited by $\mathrm{ZrTe_5}$ and paves the way for further understanding and exploration of other materials with similar properties.

Our paper is organized as follows. In Sec. II we present our computational methodology. Section III discusses the resistivity anomaly at zero magnetic field, Hall resistivity and magnetoresistance of ZrTe$_5$. Finally, section IV summarizes our work.

\section{Methedology}

One can obtain the conductivity tensor in presence of the magnetic field by solving the Boltzmann transport equation within the relaxation time approximation as\cite{ashcroft2022solid,liu2009ab},
\begin{equation}\label{equation1}
    \pmb{\sigma}^{\of{n}}\of{\boldsymbol{B}}=\frac{e^2}{\alpha \pi^3} \int d \boldsymbol{k} \tau_n \boldsymbol{v}_n(\boldsymbol{k}) \overline{\boldsymbol{v}}_n(\boldsymbol{k})\sqof{-\frac{\partial f}{\partial \varepsilon_n(\boldsymbol{k})}}, 
\end{equation}
where $\alpha$ is a spin degeneracy related number, $\alpha=4 (8)$ if spin-orbit coupling is excluded (included) in the Hamiltonian, $n$ is the band index, $f$ is the Fermi-Dirac distribution. $\varepsilon_n\of{\boldsymbol{k}}$, $\tau_n$ and $v_n\of{\boldsymbol{k}}$ are the eigenvalue, relaxation time and group velocity of the $n$-th band, respectively. $ \overline{\boldsymbol{v}}_n(\boldsymbol{k})$ describes the weighted average velocity during the past trajectory of the charge carriers ,
\begin{equation}
    \overline{\boldsymbol{v}}_n(\boldsymbol{k})=\int_{-\infty}^0 \frac{d t}{\tau_n} e^{\frac{t}{\tau_n}} \boldsymbol{v}_n\sqof{\mathbf{k}(t)}.
\end{equation}
The orbital motion of charge carriers in the reciprocal space follows the semiclassical equation of motion,
\begin{equation}\label{equation6}
    \dot{\kk}= -e\vv_n\of{\kk}\times \BB ,
\end{equation}
where the driven force of electric field was dropped off since the corresponding displacement is negligible compared to the scale of the Brillouin zone\cite{ashcroft2022solid}.

The trajectory of the charge carriers in $\boldsymbol{k}$ space is confined on a cross-section of the Fermi surface by a plane perpendicular to $\boldsymbol{B}$, since the Lorentz force does not work on charge carriers in this semiclassical framework. In order to have a more convenient and straightforward discussion of the motion of charge carriers, we define a ``scattering path length" vector as $\lll\equiv \vv(\kk)\tau$ which maps the orbitals in $\kk$ space to the real space\cite{ong1991geometric}. Consequently, assuming the magnetic field is along $\hat{z}$ axis, we rewrite the Hall conductivity in terms of the $\lll$-paths as,

\begin{equation}\label{equation7}
     \sigma_{yx}=\frac{e^3}{\of{2\pi}^3\hbar^2}\int d\varepsilon \of{-\dfrac{\partial f}{\partial\varepsilon}}\sqof{\int dk_z \mathcal{A}\of{k_z}}_{\varepsilon}B,
\end{equation}
where $\mathcal{A}(k_z)=\frac{1}{2}\of{d\lll_{\perp}\times\lll_{\perp}}\cdot \frac{\BB}{B}$ is the area swept by vector $\lll$ as the charge carriers traced out the orbitals in $\kk$ space with $\lll_{\perp}=\lll-l_z\hat{z}$\cite{ong1991geometric}. We shall consider corresponding Hall resistivity calculations in much more detail in Sec. III B.

Different from the metals with large Fermi surfaces where the carrier density does not change significantly with temperature variation, the chemical potential and concentration of charge carriers in narrow-gap semiconductors, such as Zirconium Pentatelluride ZrTe$_5$, are sensitive to the change of temperature, necessitating the precise dynamic calculations. Therefore we list our self-contained calculation procedure in the following.

\begin{enumerate}[label=(\roman*)]
\item {\it{Obtain $\rho\tau(B, \mu, T)$:}} First of all, we calculate the band-resolved magneto-conductivity over the relaxation time, denoted as $\sigma_n(\BB)/\tau_n$, on a suitable $(\mu, T)$ grid by employing Eq.(\ref{equation1}). Then we obtain the product of resistivity and relaxation time, $\rho\tau =\left[\sum_n \sigma_n(\BB)/\tau_n\right]^{-1}$. Here we assume that relaxation time for all carriers have the same temperature dependence but with different rations. For example, in ZrTe$_5$, $\tau_h(T)=5\tau_e(T)$.
\item \itt{Get $\mu(T)$:} 
We assume that the net carrier concentration, defined as $n_0=n_e-n_h$, remains constant with varying temperature. This assumption is mostly valid for semiconductors below their melting point. Consequently, the temperature dependent chemical potential $\mu(T)$, is determined by the condition:
\begin{equation}\label{equation5}
\begin{aligned}
    n_0&=n_e-n_h \\
    &=\int_{E_\text{CBM}}^{+\infty} g_c(\varepsilon) f(\varepsilon-\mu) d \varepsilon \\
    &- \int_{-\infty}^{E_\text{VBM}} g_h(\varepsilon) \sqof{1-f(\varepsilon-\mu)} d \varepsilon,
\end{aligned}
\end{equation}
where $n_e$ and $n_h$ are the concentration of electron and hole charge carriers respectively. $E_\text{CBM}$ and $E_\text{VBM}$ are the conduction band minimum and valence band maximum, respectively. $g_c(\varepsilon)$ and $g_h(\varepsilon)$ are density of states (DOS) of conduction and valence bands, respectively.
\item \itt{Interpolation to get $\rho\tau(\BB, \mu(T), T)$:} For a fixed $n_0$, corresponding to a specified doped sample or a thin film at a particular gate voltage, we can obtain the product of temperature-dependent resistivity and relaxation time, $\rho(\BB, \mu(T), T)*\tau$, by interpolating the calculated $\rho(\BB)*\tau$ on a dense grid of $(\mu, T)$ in the step (i), see Fig\ref{figs0}.
\item \itt{Fitting $\tau(T)$ with experiments:} Finally, for any given doping concentration $n_0$, we obtain the magneto-resistivity $\rho(\BB)$ at arbitrary temperature by substituting the fitted $\tau(T)$ in $\rho(\BB, \mu(T), T)*\tau$. Fitting $\tau(T)$ in semiconductors with experiment data is not as straightforward as in metals/semimetals. In metals/semimetals, the charge carrier concentration does not change significantly, allowing a consistent treatment of $\tau(T)$ across the whole temperature range. However, in semiconductor like ZrTe$_5$, which exhibits different properties with varying temperature, we need to separate the high temperature regime from the low one. The metallic behavior of ZrTe$_5$ at low temperature allows us to fit $\tau(T)$ with corresponding experiment results in this regime. On the contrary, the semiconductor behavior in the high temperature regime precludes the direct derivation of $\tau(T)$ from experimental measurements. Nonetheless, it is well-known that the electron-phonon scattering described by $\tau(T) \propto 1/T$ dominates at the high temperature regime. We assume $\tau(T)\propto \rho_{sc}(T)$, and $\rho_{sc}(T)$ represents the scattering-related resistivity which is comprised in the relaxation time. The distinct scattering behavior of ZrTe$_5$ across different temperature regime could be depicted jointly by the Bloch-Gr\"uneisen model\cite{ziman, liu2023gate} as,

\footnotesize{
\begin{equation}\label{equation8}
    \rho_{sc}(T)=\rho_0+\alpha\left(\frac{T}{\Theta_R}\right)^n \int_0^{\frac{\Theta_R}{T}} \frac{x^n}{\left(e^x-1\right)\left(1-e^{-x}\right)} d x,
\end{equation}}
    \normalsize
with the parameters $\rho_0=1.06$, $\alpha=11$, $n=2$, and $\Theta_R=600$ obtained by fitting the experiment data\cite{liu2023gate} at low temperature. $\tau(T)$ in this work are calculated by Eq.(\ref{equation8}) and the hypothesis $\tau(T)\propto \rho_{sc}(T)$.

\end{enumerate}

\begin{figure*}[htbp]
    \centering 
    \includegraphics[width=0.95\textwidth]{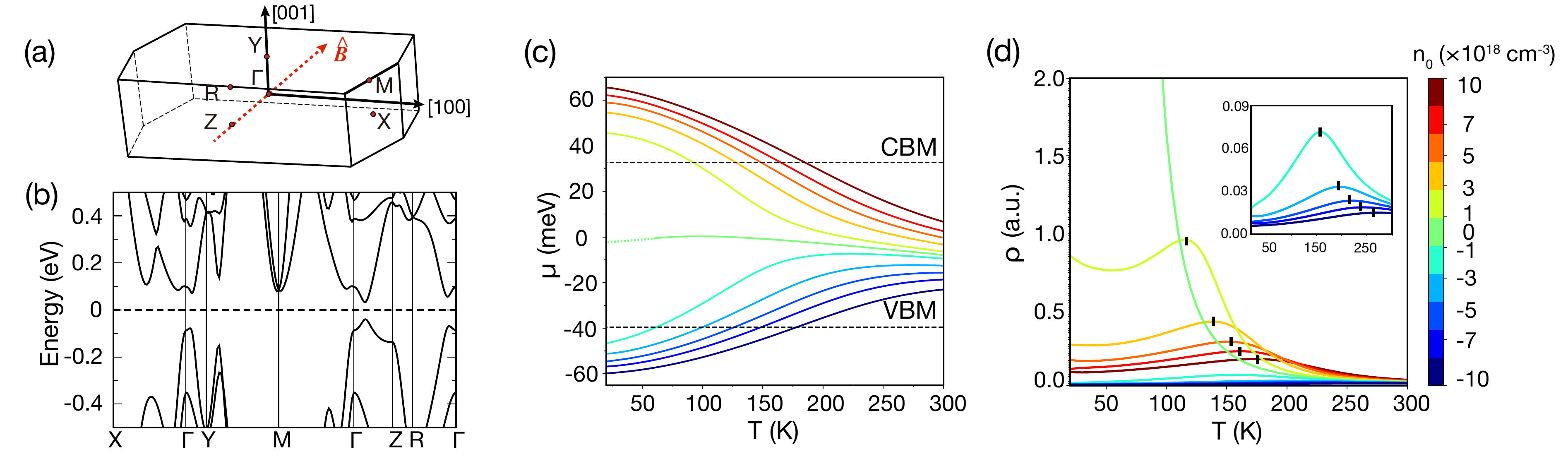}
    \caption{(a) The Brillouin zone of $\mathrm{ZrTe_5}$. The applied current and magnetic field are parallel to the crystallographic $a$ and $b$ axis, respectively. The orientation of magnetic field is denoted as $z$ axis throughout the text. (b) The band structure of $\mathrm{ZrTe_5}$. (c) The temperature-dependent chemical potential with different carrier densities $n_0$ indicated by different colors. $\mathrm{ZrTe_5}$ with $n_0 >0$ is electron-doped while $n_0<0$ means hole-doped $\mathrm{ZrTe_5}$.The high and low horizontal black dash line denote the valence band maximum (VBM) = -40 meV and conduction band maximum (CBM) = 33 meV, respectively. The green dash line is the extrapolated chemical potential of $n_0$ = 0 cm$^{-3}$ at $T < 70\ K$ which can not be well determined with non-zero energy gap. (d) The temperature-dependent zero-field resistivity with different carrier densities. The inset shows the resistivity of hole-doped $\mathrm{ZrTe_5}$. The small black square denotes the peak of resistivity. We use a color bar to indicate the relation between temperatures and doping densities.\label{fig1}}
\end{figure*}
Note that this procedure for calculating magneto-resistivity is not limited to a special compound ZrTe$_5$. This approach is applicable to a broad range of narrow-gap semiconductors. The band structure was calculated using the Vienna ab initio simulation package (VASP)\cite{kresse1996efficient} with the generalized gradient approximation of Perdew, Burke, and Ernzerhof for the exchange-correlation potential\cite{perdew1996generalized}. We performed the self-consistent calculation on a k-mesh of 11×11×11 with energy cutoff of 550 eV. The temperature-dependent chemical potential was calculated using BoltzTrap\cite{madsen2006boltztrap} package and performed on the k-mesh of 50×50×25. Magneto-conductivity and Fermi surface were calculated using the WANNIERTOOLS package\cite{wu2018wanniertools} based on a tight-binding model constructed by the WANNIER90 package\cite{marzari1997maximally,souza2001maximally,marzari2012maximally,mostofi2014updated}. In the Magneto-conductivity calculation, we use a k-mesh of 201×201×201 and set the broadening width of Fermi-Dirac distribution function to 200 meV.

\section{Results and Discussions}

\subsection{Resistivity anomaly}\label{sectionI}

ZrTe$_5$, a typical transition-metal pentatelluride compound and a representative narrow-gap semiconductor, has been extensively studied due to its prominent thermoelectric properties. Before subjecting it to the magnetic field, we shall first study its temperature-dependent resistivity at zero field due to its inherent thermal sensitivity as a semiconductor. We present the calculated band structure of ZrTe$_5$ and the corresponding high symmetry path in the Brillouin zone in Fig.\ref{fig1} (a-b). It is shown in Fig.\ref{fig1} (b) that an indirect narrow gap of about 73 $\rm{meV}$ occurs along the $\Gamma$-Z direction with strong electron-hole asymmetry. 

Because the transport behavior is highly dependent on the doping levels\cite{liu2023gate}, we calculated the temperature-dependent chemical potential $\mu(T)$ with a series of doping levels, as shown in Fig.\ref{fig1} (c). The doping level is denoted as $n_0 = n_e-n_h$, where positive/negative values represent electron/hole doping systems. The chemical potentials move towards the middle of the energy gap with increasing temperature, irrespective of the type of doping. This agrees qualitatively with the ARPES observation\cite{zhang2017electronic} and results from the intrinsic thermodynamics of materials that the chemical potential favors lower DOS with increasing temperatures\cite{colinge}.  Note that when the doping level is fixed as $n_0 = 0$, the chemical potential is ill-defined at low temperature. Therefore we only sketch the tendency of $\mu$ at low temperature relying on the results at higher temperatures, as shown by the green dashed line segment in Fig.\ref{fig1} (c). 

Furthermore, we calculate the temperature-dependent resistivity at different doping levels as shown in Fig.\ref{fig1} (d). The resistivities of all doped systems (excluding the charge neutrality system) increase at low temperatures and then drop quickly after attaining their peaks at critical temperatures $T_p$. In order to discuss the origin of the resistivity anomaly, we introduce two important variables: the total charge carrier concentration $n_t=n_e+n_h$, which might vary with temperature due to thermal excitation, and the doping level $n_0=n_e-n_h$, assumed as a constant. At low temperatures, where the thermal energy is insufficient to excite electrons from valence bands to conduction bands, $n_t$ remains nearly unchanged. Because the relaxation time decreases with temperature, the resistivity increases and ZrTe$_5$ behaves like a metal. However, as the temperature rises sufficiently to excite electrons from valence bands to conduction bands and generate electron-hole pairs, $n_t$ increases and  ZrTe$_5$ behaves like a semiconductor. 

The contradicted dependence on temperature of relaxation time and total charge carrier concentration implies the existence of a critical temperature $T_P$, at which the resistivity stops increasing with temperature. To further investigate the resistivity anomaly, we mark resistivity peaks at $T_P$ with black squares in Fig.\ref{fig1} (d). We found that $T_P$ grows with the magnitude of $n_0$. This behavior is due to the higher temperature required to generate thermal electron-hole pairs in systems with larger doping densities. Moreover, given that ZrTe$_5$ has a significant particle-hole asymmetry, $T_P$ of electron-doped is lower than that of the hole-doped with the same magnitude of $n_0$, owing to the larger DOS of valence bands\cite{liu2023gate,shahi2018bipolar}. Similar multi-carrier and thermally-excited-carrier theories have been proposed in previous works\cite{liu2023gate,shahi2018bipolar,fu2020dirac,wang2021thermodynamically}.

\subsection{Sign reversal of Hall resistivity}

Until now, we have been only concerned with the resistivity in the absence of magnetic field. In the following, we are going to investigate the magnetotransport properties. We shall employ the method introduced in Sec.II. Before proceed to the concrete properties, we explain Eq.(\ref{equation7}) in detail. The transverse conductivity $\sigma_{xy}$ could be expressed in terms of the area $\mathcal{A}\of{k_z}$ swept by $\lll$ (see appendix for detailed derivation of Eq.(\ref{equation7})). The trajectory orbits in $\kk$ space can be divided into two categories. One is the orbits only composed of convex part, which is transformed to a simple $\lll$ path with single loop and a certain circulation orientation. Thus the corresponding Hall resistivity exhibits the expected sign. The other one is the orbits with concave segments, which is converted into intersected $\lll$ path with second loops adjacent to the primary one and hence might change circulation orientations of the whole (or part of) orbit. Whether and to what extent the circulation orientation of the orbit changed depends on the fine curvature of the $\kk$ orbits and decides the final sign of Hall conductivity of this single orbit \cite{ong1991geometric}. 

 \begin{figure*}[!htp]
    \centering 
    \includegraphics[width=0.95\textwidth]{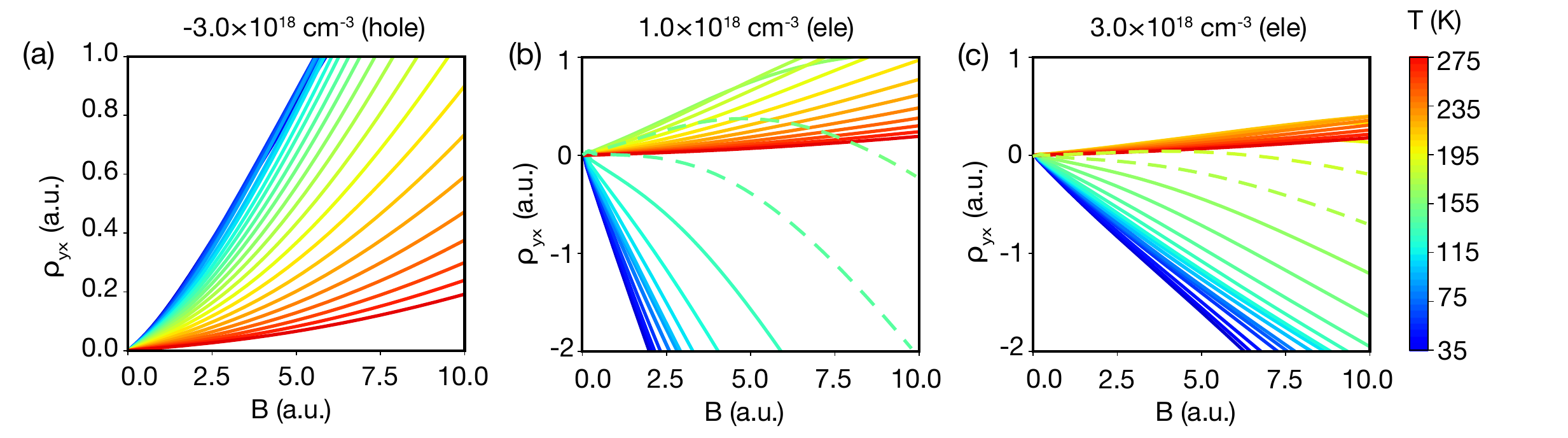}
    \caption{ The field-dependent Hall resistivity at different temperatures with the doping densities of (a) $n_0 = -3.0\times 10^{18}\ \text{cm}^{-3}$, (b) $1.0\times 10^{18}\ \text{cm}^{-3}$, and (c) $3.0\times 10^{18}\ \text{cm}^{-3}$, respectively. The dashed lines indicate the Hall resistivities that reverse sign with the increasing magnetic field. Temperature is varied from 35 K to 275 K at steps of 10 K. The relation between the temperature and color is indicated in the color bar. \label{fig2}}
\end{figure*}
When electron and hole charge carriers both contribute to electric current, the competition between them is crucial to diagnose the behavior of $\rho_{xy}$, such as the nonlinear or sign reversal features. Intuitively, the charge carrier with a larger concentration dominates the transport. However, it only holds under the condition of sufficiently strong magnetic fields. At the low magnetic field, charge carriers only trace part of the orbit, the specific area $\mathcal{A}\of{k_z}$ swept by $\lll$ sensitively depends on the fine structure of the Fermi surface and orientation of the magnetic field. Accordingly, we classify the characteristic quantities of the cyclotron movement $\omega_c\tau = \frac{e B \tau}{m^*}$ ($m^*$ is the cyclotron mass) into two limiting cases, i.e., $\omega_c\tau\ll 1$ (low-field limit) and $\omega_c\tau\gg 1$(high-field limit), where the magnetoresistance (MR) and Hall resistivity might exhibit distinct features as shown in the following.

We begin with the Hall resistivity of ZrTe$_5$ with doping level $n_0 = -3.0\times 10^{18}\ \text{cm}^{-3}$ (hole-doped), $1.0\times 10^{18}\ \text{cm}^{-3}$ (electron-doped), and $3.0\times 10^{18}\ \text{cm}^{-3}$ (electron-doped), which are shown in Fig.\ref{fig2}. The shape of the Hall resistivity varies drastically with increasing temperature. However, there is an obvious difference between the Hall resistivity of the hole and electron-doped system. The Hall resistivity curves for hole-doped in Fig.\ref{fig2} (a) only exhibit nonlinear feature , while the ones for electron-doped in Fig.\ref{fig2} (b,c) show both sign reversal and nonlinear features. At low temperatures where there is only a single type of charge carrier, the magnitude of Hall resistivity monotonously increases with the magnetic field and the system shows the expected Hall sign, i.e., positive/negative for hole/electron-doped systems. With increasing temperature, the magnitude of Hall resistivity $\rho_{xy}$ drops for both electron and hole-doped systems, due to the thermal excitation and the increasing charge carriers contributing to the transport process. When reaching the intermediate temperature (around 155K), the Hall resistivity of the electron-doped system undergoes a sign reversal with a strong nonlinear slope as shown in Fig.\ref{fig2} (b,c). On the contrary, the Hall resistivity of the hole-doped system only exhibits nonlinear features without sign reversal as temperature rises. 

\begin{figure*}[!htbp]
    \centering 
    \includegraphics[width=0.95\textwidth]{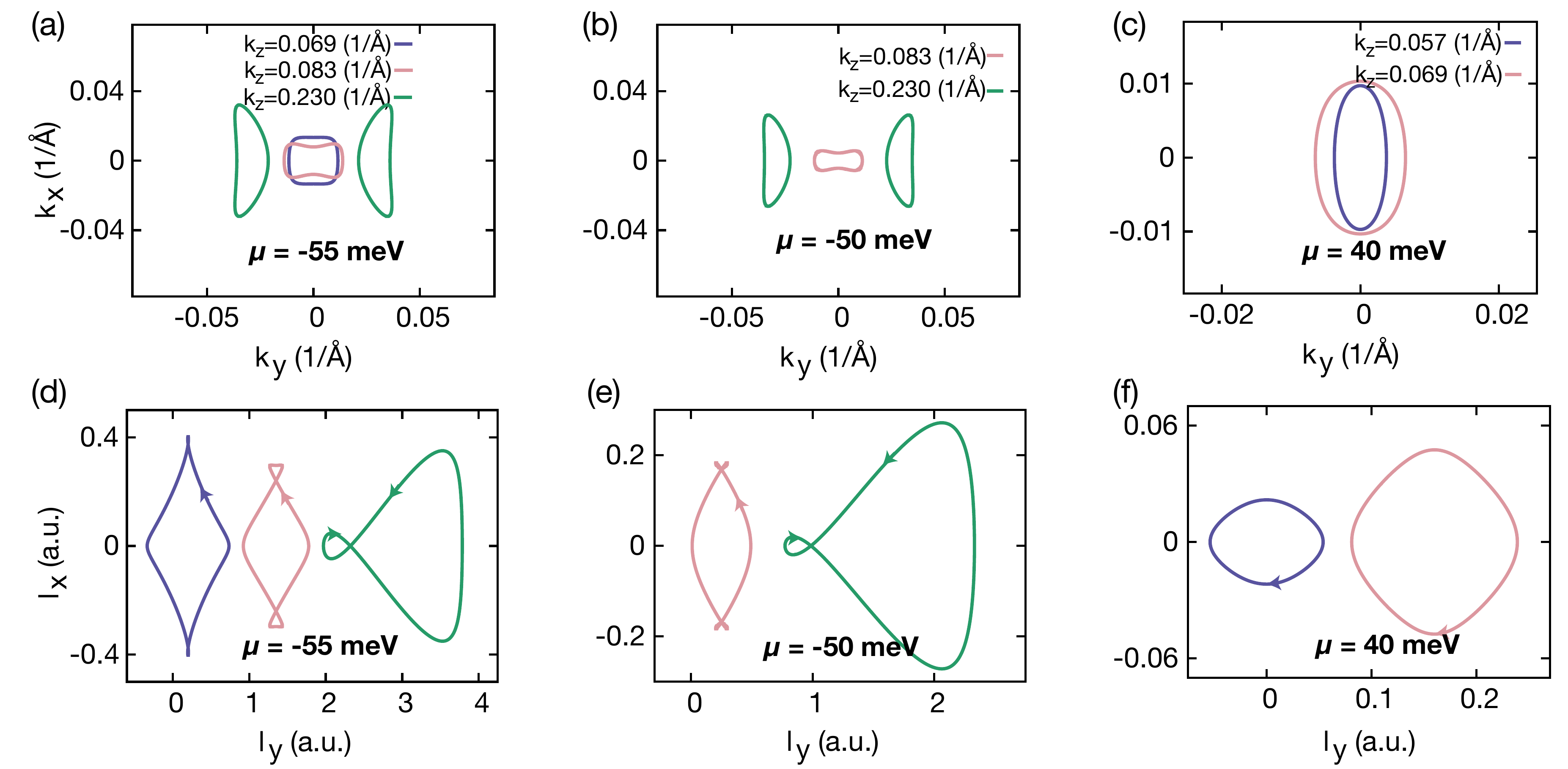}
    \caption{ (a-c) The typical cross section of Fermi surfaces and (d-f) their corresponding $l$-paths with the magnetic field along $z$ direction. The chemical potentials are set to (a,d) $\mu=-55$ meV, (b,e) $\mu=-50$ meV, and (c,f) $\mu=40$ meV, respectively.   \label{fig3}}
\end{figure*}

According to our analysis of $\sigma_{xy}$, the nonlinear and sign reversal features of Hall resistivity probably stem from the complicated competition between electron and hole carriers (including both intrinsic and thermally excited ones). In order to verify this,  we plot representative calculated $\kk$-orbits and its corresponding $\lll$-paths at different chemical potentials $\mu_0 = -55, -50$ and $ 40 \rm \ meV$ in Fig.\ref{fig3}. All the representative $\kk$-orbits in the hole-doped system have concave segments (Fig.\ref{fig3}(a-b)), which gives rise to the intersected $\lll$-path with a tiny secondary loop and the opposite circulation. It means a small part of charge carriers on these hole pockets behave like electrons, resulting in the slight nonlinear feature of corresponding Hall resistivity (Fig.\ref{fig2}(a)). However, the electron pockets are all spherical surfaces and the corresponding $\lll$-paths are simple as shown in Fig.\ref{fig3}(c, f), indicating linear Hall resistivity in electron-doped systems at low temperatures. 

When temperature increases, the sign reversal at low magnetic field only happens in the electron-doped systems, which is difficult to explain by the qualitative orbits analysis. Therefore we employ the two-band model, where the Hall resistivity is written as (add ref when submit), 

\begin{align}
\rho_{yx}=\frac{1}{e} \frac{(n_h\mu_h^2-n_e\mu_e^2)B+(n_h-n_e)\mu_e^2\mu_h^2 B^3}{(n_e\mu_e+n_h\mu_h)^2+
    (n_h-n_e)^2\mu_e^2\mu_h^2B^2}\label{eqn:hallB0} 
\end{align}
where $n_{e/h}$, $\mu_{e/h}$ are the concentration and mobility of electron/hole carriers, respectively. With the fixed doping level $n_0 = n_e -n_h$, we use $n_e=n_e^0 + \delta_n, n_h = \delta_n $ for electron-doped, and $n_h=n_h^0 + \delta_n, n_e = \delta_n $ for hole-doped in Eq.(\ref{eqn:hallB0})  as following,

\begin{align}
\rho_{yx}^{hole}=\frac{1}{e} \frac{ \{ [ \delta_n (\mu_h^2-\mu_e^2) + n_h^0\mu_h^2] + n_h^0 \mu_e^2\mu_h^2 B^2 \} B}{(n_e\mu_e+n_h\mu_h)^2+
    (n_h-n_e)^2\mu_e^2\mu_h^2B^2}\label{hole-dopped} 
\end{align}

\begin{align}
\rho_{yx}^{ele}=\frac{1}{e} \frac{ \{ [ \delta_n (\mu_h^2-\mu_e^2) - n_e^0\mu_e^2] - n_e^0 \mu_e^2\mu_h^2 B^2 \} B}{(n_e\mu_e+n_h\mu_h)^2+
    (n_h-n_e)^2\mu_e^2\mu_h^2B^2}\label{electron-dopped} 
\end{align}
where $n_e^0$ and $n_h^0$ are the concentration of electron and hole carriers at zero-temperature. Because the mobility of hole carrier is larger than that of electron in ZrTe$_5$\cite{liu2023gate,shahi2018bipolar}, the Hall resistivity $\rho_{yx}^{hole}$ is always positive due to the positive terms, $n_h^0 \mu_h^2$ and $\delta_n(\mu_h^2-\mu_e^2)$, in Eq.(\ref{hole-dopped}). However, the numerator of electron-doped $\rho_{yx}$ in Eq.(\ref{electron-dopped}) are firstly negative due to the term $-n_e^0 \mu_e^2 B - n_e^0 \mu_e^2\mu_h^2 B^3 $ and then might change sign because of the growing amount of excited carriers $\delta_n $ with increasing temperature. 

Furthermore, we can employ Eq.(\ref{electron-dopped}) to understand the sign change of electron-doped Hall resistivity with increasing magnetic field at intermediate temperatures (see green dashed curves in Fig.\ref{fig2}(b-c)). Supposing the Hall resistivity at weak magnetic field is positive, the numerator of Eq.(\ref{electron-dopped}) satisfies the condition that $\{ [ \delta_n (\mu_h^2-\mu_e^2) - n_e^0\mu_e^2] - n_e^0 \mu_e^2\mu_h^2 B^2 \} > 0$ when $\omega_c \tau \ll 1 $. With the increasing magnetic field, the second term of the numerator $- n_e^0 \mu_e^2\mu_h^2 B^2$ will increase and leads to the negative Hall resistivity at critical $\BB$ field. Nevertheless, if the temperature rises at the same time, $\delta_n$ would increase and compensate the effect induced by a stronger magnetic field. It suggests the occurrence of sign reversal necessitates a higher magnetic field under elevated temperatures. This phenomenon is displayed in Fig.\ref{fig2}(b,c), where we show the Hall resistivity with sign reversal by dashed lines. It is noteworthy that at the higher temperature (indicated by the curves in reds), the magnitude of $\BB$ field required for sign reversal is far beyond our calculation and experimental measurement scope. 

Although the understanding based on the two-band model is rudimentary, the physical mechanism in the above discussion, i.e., competition between electron and hole carriers driven by the magnetic field and thermal excitation, is the key origin of the nonlinear and sign reversal features of the Hall resistivity. Furthermore, by using Eq.(\ref{equation7}) to analyze the Hall resistivity in electron/hole-doped systems under low fields ($\omega_c \tau \ll 1$) and the well-established fact that charge carriers with the larger concentration dominate under high fields ($\omega_c \tau \gg 1$), we arrive at the same conclusion. To avoid redundancy, we will not not give detailed repetitive explanations.

\subsection{Nonsaturating magnetoresistance}
\begin{figure*}[htbp]
    \centering 
    \includegraphics[width=0.95\textwidth]{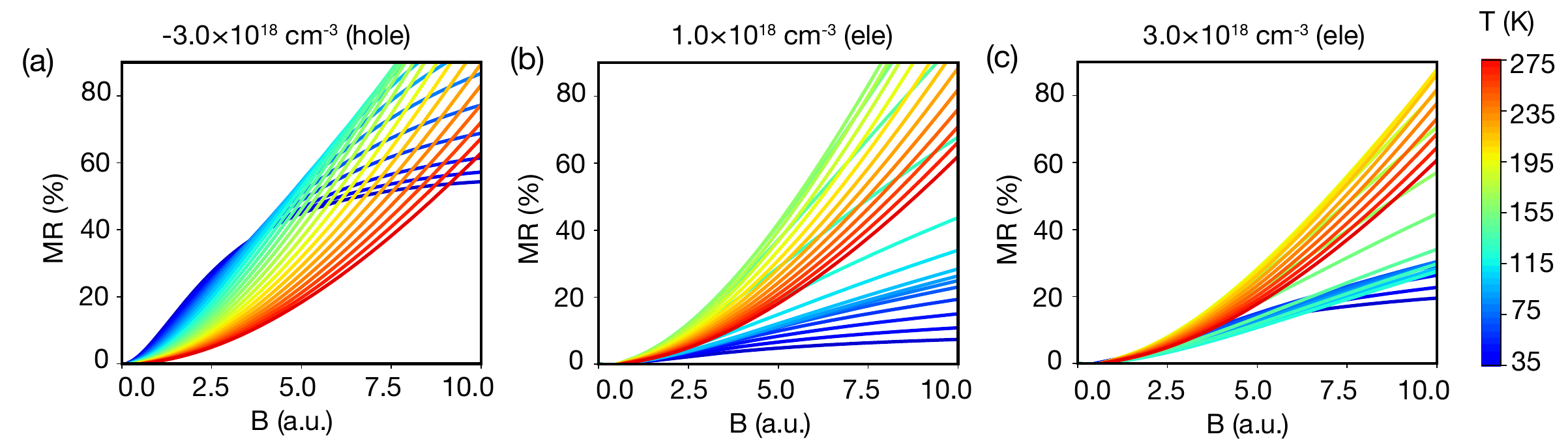}
    \caption{  The field-dependent MR at different temperatures with the fixed charge carrier concentration of (a) $n_0 = -3.0\times 10^{18}\ \text{cm}^{-3}$, (b) $1.0\times 10^{18}\ \text{cm}^{-3}$, and (c) $3.0\times 10^{18}\ \text{cm}^{-3}$, respectively. Temperature is varied from 35 K to 275 K at steps of 10 K. The relation between the temperature and color is indicated in the color bar. \label{fig4}}
\end{figure*}

To proceed with investigating the MR of ZrTe$_5$, we plot our calculated MR results of the same doping level as the Hall resistivity in Fig.\ref{fig4}. It's apparent that at high temperatures, the MR of all doped levels exhibit non-saturating behavior. This phenomenon is due to the presence of both electron and hole carriers in those doped systems because of the thermal excitation. The nearly charge carrier compensation leads to the consistent rise in MR with the magnetic field, which agrees well with the experimental measurement\cite{liu2023gate,lozano2022anomalous}. 

In the low-temperature range, the MR of electron and hole doped cases both show the trend toward saturation. However, the details of saturation are slightly different for electron and hole-doped systems. For the hole-doped case as shown in Fig.\ref{fig4} (a), the MR saturates more slowly and at a larger magnetic field than that in electron-doped systems Fig.\ref{fig4} (b,c) (see curves in blues). We again refer to Fig.\ref{fig3} to explain the distinction between these two systems. At low temperatures where there are seldom excited charge carriers, the intrinsic ones dominate the transport. Regarding the fact that the hole pockets with concave segments give rise to both electron and hole-like charge carriers, the slightly compensation between them postpones the saturation of MR. On the contrary, because the electron pockets possess only spherical surfaces, electron-doped systems exhibit rapid saturation of MR as expected. Compared with the experiments\cite{liu2023gate}, our results does not show the negative MR and quantum oscillation at low temperatures. Moreover, the experimental MR saturation at low temperatures is more obvious in electron-doped ZrTe$_5$. Here we must point out that our calculated results is constricted in the semiclassical Boltzmann transport framework, i.e., without taking the Landau level, weak localization, and possible Berry curvature into account. For simplicity, we also ignore the $\kk$-dependence of the relaxation time $\tau$, which could also be a possible reason for deviation of our calculation from the experimental measurement.

\section{Conclusion}

To conclude, we developed an effective methodology to study the magnetoresistance and Hall resistivity of narrow-gap semiconductors based on the combination of first principles method and semiclassical Boltzmann transport theory.  This approach was applied to study the temperature-dependent galvanomagnetic properties of ZrTe$_5$, taking into account variations in chemical potential and charge carrier concentrations. Our calculated results successfully reproduce observed transport behaviors in experiments, such as the resistivity anomaly, sign reversal and nonlinear Hall resistivity of electron-doped ZrTe$_5$, and non-saturating magnetoresistance of hole-doped ZrTe$_5$. Our analysis domenstrates that these transport anomalies can be explained in terms of multi-carrier behavior and Fermi surface geometry. 

Although our method effectively accounts for most of the experimental observations, certain detailed characteristics such as quantum oscillations and negative magnetoresistance were not captured due to the omission of Landau levels, Berry curvature, and the influence interactions on the $k$-dependence of relaxation time. Nevertheless, our approach allows us to identify properties that can be explained by semiclassical contributions and highlights nontrivial features in ZrTe$_5$ for future investigations. Moreover, our methodology can be extended to other narrow-gap semiconductors. Although theoretical models can explain some anomalous transport properties, first principles calculations remain a powerful tool for gaining deeper insights into galvanomagnetic behavior, thereby paving ways for further exploration.

\begin{acknowledgments}
This work was supported by the National Key R\&D Program of China (Grant No. 2023YFA1607401, 2022YFA1403800), the National Natural Science Foundation of China (Grant No.12274436, 11925408, 11921004, 12174439), the Science Center of the National Natural Science Foundation of China (Grant No. 12188101) and  H.W. acknowledge support from the Informatization Plan of the Chinese Academy of Sciences (CASWX2021SF-0102). 

\end{acknowledgments}

\appendix
\section{Electric conductivity in weak magnetic field}\label{appendix A}
With the relaxation time approximation and the condition of steady state, $\dfrac{\partial g\of{\rr,\kk}}{\partial t} = 0$, BTE is expressed as
\begin{equation}
    \dfrac{dg(\rr,  \kk)}{dt}=\dfrac{\partial g(\rr, \kk)}{\partial \rr}\dfrac{d\rr}{dt}+ \dfrac{\partial g(\rr, \kk)}{\partial \kk}\dfrac{d\kk}{dt}=-\dfrac{\delta g(\rr, \boldsymbol{k})}{\tau}.
\end{equation}
$g\of{\rr,\kk}$ is the distribution function of particles in phase space. Using the semiclassical equation of motion under the electric and magnetic field,
\begin{equation}
\begin{aligned}
    \frac{d\boldsymbol{r}}{dt}&\equiv\boldsymbol{v}(\boldsymbol{k})=\frac{1}{\hbar}\nabla_{\boldsymbol{k}} \varepsilon(\boldsymbol{k})\\
    \frac{d\boldsymbol{k}}{dt}&=-\frac{e}{\hbar}\sqof{\boldsymbol{E}+\vv\of{\kk}\times\BB}
\end{aligned}
\end{equation}
we rewrite BTE as
\begin{equation}\label{equation2}
\begin{aligned}
        &\dfrac{\partial g(\rr, \kk)}{\partial \rr}\cdot\vv(\kk)+ \dfrac{\partial g(\rr, \kk)}{\partial \kk}\cdot\cuof{-\frac{e}{\hbar}\sqof{\boldsymbol{E}+\vv\of{\kk}\times\BB}}\\
        =&-\dfrac{\delta g(\rr, \boldsymbol{k})}{\tau}
\end{aligned}
\end{equation}
When the applied fields are weak, $g\of{\rr,\kk}$ can be written as 
\begin{equation}
    g\of{\rr,\kk} = f\of{\rr,\kk}+\delta g\of{\rr,\kk},
\end{equation}
where $f\of{\rr,\kk}$ is the equilibrium Fermi-Dirac function and $\delta g\of{\rr,\kk}$ describes the deviation from equilibrium. Because $\delta g\of{\rr,\kk}$ is a small quantity, we can replace $g\of{\rr,\kk}$ with $f\of{\rr,\kk}$ in Eq.(\ref{equation2}),
\begin{equation}
\begin{aligned}
    &\dfrac{\partial f(\rr, \kk)}{\partial \rr}\cdot\vv\of{\kk}
    +\dfrac{\partial f(\rr, \kk)}{\partial \kk}\cdot\cuof{-\frac{e}{\hbar}\sqof{\boldsymbol{E}+\vv\of{\kk}\times\BB}}\\
    =&-\dfrac{\delta g(\rr,\kk)}{\tau}
\end{aligned}
\end{equation}
As the $\rr$-dependence and $\kk$-dependence of $f\of{\rr,\kk}$ is through $f\sqof{\mu\of{\rr},T\of{\rr},\varepsilon(\kk)}$, we have
\begin{equation}\label{equation3}
\begin{aligned}
    &\vv\cdot \cuof{
    \frac{\partial f}{\partial T}\Bnabla_{\rr} T+
    \frac{\partial f}{\partial \mu}\Bnabla_{\rr}\mu
    -e\frac{\partial f}{\partial \varepsilon}\sqof{\boldsymbol{E}+\vv\times\BB}}\\
    =&-\dfrac{\delta g}{\tau}
\end{aligned}
\end{equation}
For convenience, we suppress explicit reference to $\rr$ and $\kk$. 

As $\vv\cdot\of{\vv\times\BB}=0$, we can see that the replacement of $g$ with $f$ in Eq.(\ref{equation2}) leads to the vanish of $\BB$. Therefore, we only replace $g$ with $f$ in terms that do not contain $\BB$ and modify Eq.(\ref{equation3}) as
\begin{equation}
\begin{aligned}
    -\dfrac{\delta g}{\tau}=
    &\vv\cdot \of{
    \frac{\partial f}{\partial T}\Bnabla_{\rr} T+
    \frac{\partial f}{\partial \mu}\Bnabla_{\rr}\mu
    -e\frac{\partial f}{\partial \varepsilon}\EE}\\
    -&\frac{e}{\hbar}\of{\vv\times\BB}\cdot\Bnabla_{\kk}\of{\delta g}
\end{aligned}
\end{equation}
Assuming the absence of a temperature gradient field and the external electric and magnetic fields being spatially uniform, we can obtain the expression of $\delta g$ from the above equation,
\begin{equation}
\begin{aligned}
    \delta g=&\of{1+\tau\Omega}^{-1}\of{\dfrac{\partial f}{\partial \varepsilon}}e\vv\tau\cdot \EE,
\end{aligned}
\end{equation}
where $\Omega=-\dfrac{e}{\hbar}\of{\vv\times\BB}\cdot\Bnabla_{\kk}$. Using Jones–Zener expansion,
\begin{equation}
    (1+\tau \Omega)^{-1}=1-\tau \Omega+(\tau \Omega)^2-\ldots
\end{equation}
and keeping only the first order of $\BB$, We have
\begin{equation}
\begin{aligned}
    \delta g\of{\rr,\kk}=&\of{1-\tau \Omega}\of{\dfrac{\partial f}{\partial \varepsilon}}e\vv\tau\cdot \EE,
\end{aligned}
\end{equation}
and the induced electric current,
\begin{equation}
\begin{aligned}
    \JJ=&\int \frac{d\kk^3}{\of{2\pi}^3}\of{-\dfrac{\partial f}{\partial \varepsilon}}e^2\vv\tau\of{\vv\cdot \EE}\\
    -&\int \frac{d\kk^3}{\of{2\pi}^3}\of{-\dfrac{\partial f}{\partial \varepsilon}}e^2\vv\tau\Omega\of{\vv\tau\cdot \EE}.
\end{aligned}
\end{equation}
The first term describes the electrical current irrelevant of the magnetic field, and the second term describes the Hall current. 

Assuming $\BB=B\hat{z}$, the Hall conductivity is
\begin{equation}
\sigma_{yx} = \frac{e^3}{\hbar}\int \frac{d\kk^3}{\of{2\pi}^3}\of{-\dfrac{\partial f}{\partial \varepsilon}}\of{v_y\tau}\of{\vv\times\BB}\cdot\Bnabla_{\kk}\of{v_x\tau}.
\end{equation}
If $S_{\kk}$ and $\kk_{n}$ denote the iso-energy surface and the normal vector of the iso-energy surface, respectively, we can obtain the following equation at low temperature, $T\ll \varepsilon_F$\cite{ong1991geometric},
\begin{equation}
\begin{aligned}
     \int d\kk^3\of{-\dfrac{\partial f}{\partial \varepsilon}}
     &=\int dk_{n}dS_{\kk} \of{-\dfrac{\partial f}{\partial \varepsilon}}\\
     &= \int \frac{d\varepsilon}{\hbar v}dS_{\kk}\of{-\dfrac{\partial f}{\partial \varepsilon}}\\
     &=\int_{\varepsilon_{F}}\frac{dS_{\kk}}{\hbar v} ,
\end{aligned}
\end{equation}
where $\varepsilon_F$ is the Fermi energy. Thus, the integration over reciprocal space reduces to the integral over the Fermi surface, 
\begin{equation}\label{equation4}
\sigma_{yx}=\frac{e^3}{\of{2\pi}^3\hbar^2}\int_{\varepsilon_{F}}\frac{dS_{\kk}}{ v}\of{v_y\tau}\of{\vv\times\BB}\cdot\Bnabla_{\kk}\of{v_x\tau}.
\end{equation}
We denote the tangential vector of the electron orbital on Fermi surface as $dk_t m$, and the tangential vector of Fermi surface that is perpendicular to $k_t$ as $k_a$. Therefore,
$dS_{\kk}=dk_tdk_a=dk_tdk_z\dfrac{v}{v_\perp}$, where $v_{\perp}$ is the component of $v_n$ normal to $B_z$, and Eq.(\ref{equation4}) can be simplified as
\begin{equation}\label{equation15}
\begin{aligned}
    \sigma_{yx}&=\frac{e^3}{\of{2\pi}^3\hbar^2}\int_{\varepsilon_{F}} dk_z\oint dk_t\of{v_y\tau}\of{\frac{\vv}{v_{\perp}}\times\BB}\cdot\Bnabla_{\kk}\of{v_x\tau}\\
    &=\frac{e^3}{\of{2\pi}^3\hbar^2}\int_{\varepsilon_{F}} dk_z\oint dk_tl_y\sqof{B\hat{t}\cdot\Bnabla_{\kk}\of{l_x}}\\
    &=\frac{e^3}{\of{2\pi}^3\hbar^2}\int_{\varepsilon_{F}} dk_z\int dl_xl_yB\\
    &=\frac{e^3}{\of{2\pi}^3\hbar^2}\int_{\varepsilon_{F}} dk_z \mathcal{A}\of{k_z}B
\end{aligned}
\end{equation}
We define $\lll\equiv \vv\tau$ as the ``scattering path length" vector in the second row, $\mathcal{A}(k_z)=\frac{1}{2}\of{d\lll_{\perp}\times\lll}_{\perp}\cdot \frac{\BB}{B}$  is the area swept by $\lll$ in ``$\lll$-space" when the charge carriers move along the orbitals at $k_z$, and $\lll_{\perp}=\lll-\lll\cdot\hat{z}$\cite{ong1991geometric}. At higher temperatures, Eq.(\ref{equation15}) should be modified as
\begin{equation}
    \sigma_{yx}=\frac{e^3}{\of{2\pi}^3\hbar^2}\int d\varepsilon \of{-\dfrac{\partial f}{\partial\varepsilon}}\sqof{\int dk_z \mathcal{A}\of{k_z}}_{\varepsilon}B.
\end{equation}
Here, we integrate the charge carrier orbitals on iso-energy surface $S\of{\varepsilon}$ weighted by $-\frac{\partial f}{\partial\varepsilon}$.

	\setcounter{figure}{0}
	\renewcommand{\thefigure}{S\arabic{figure}}%
 \begin{figure*}[htbp]
    \centering 
    \includegraphics[width=0.7\textwidth]{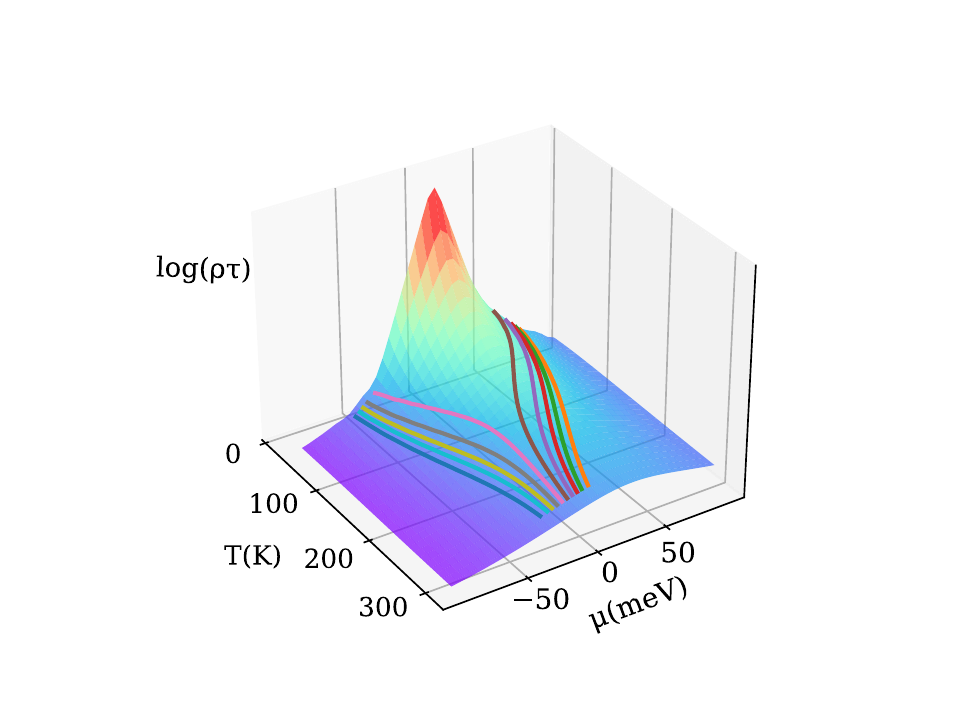}
    \caption{ Schematic illustration of interpolation process. The two-dimensional surface represents calculated $\mathrm{log}\of{\rho\tau}$ on the grid $(\mu, T)$, and lines with different colors indicate $\rho(\BB,\mu(T),T)*\tau$ for different concentrations $n_0$.   \label{figs0}}
\end{figure*}
\begin{figure*}[htbp]
    \centering 
    \includegraphics[width=0.7\textwidth]{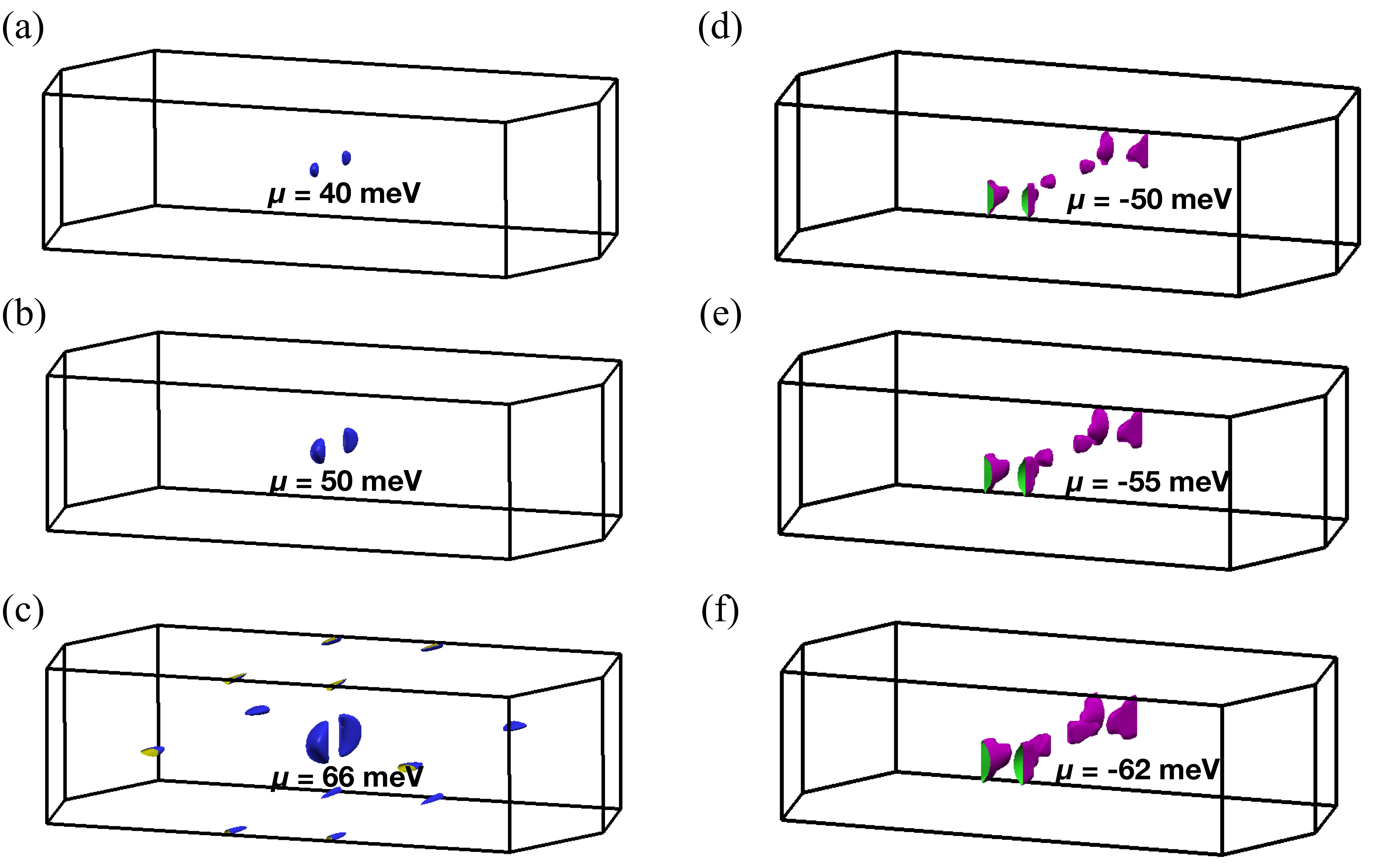}
    \caption{  The Fermi surfaces with the chemical potential $\mu=$ (a) 40 meV, (b) 50 meV, (c) 66 meV, (d) -50 meV, (e) -55 meV, (f) -62 meV.  \label{figs1}}
\end{figure*}

\begin{figure*}[htbp]
    \centering 
    \includegraphics[width=0.95\textwidth]{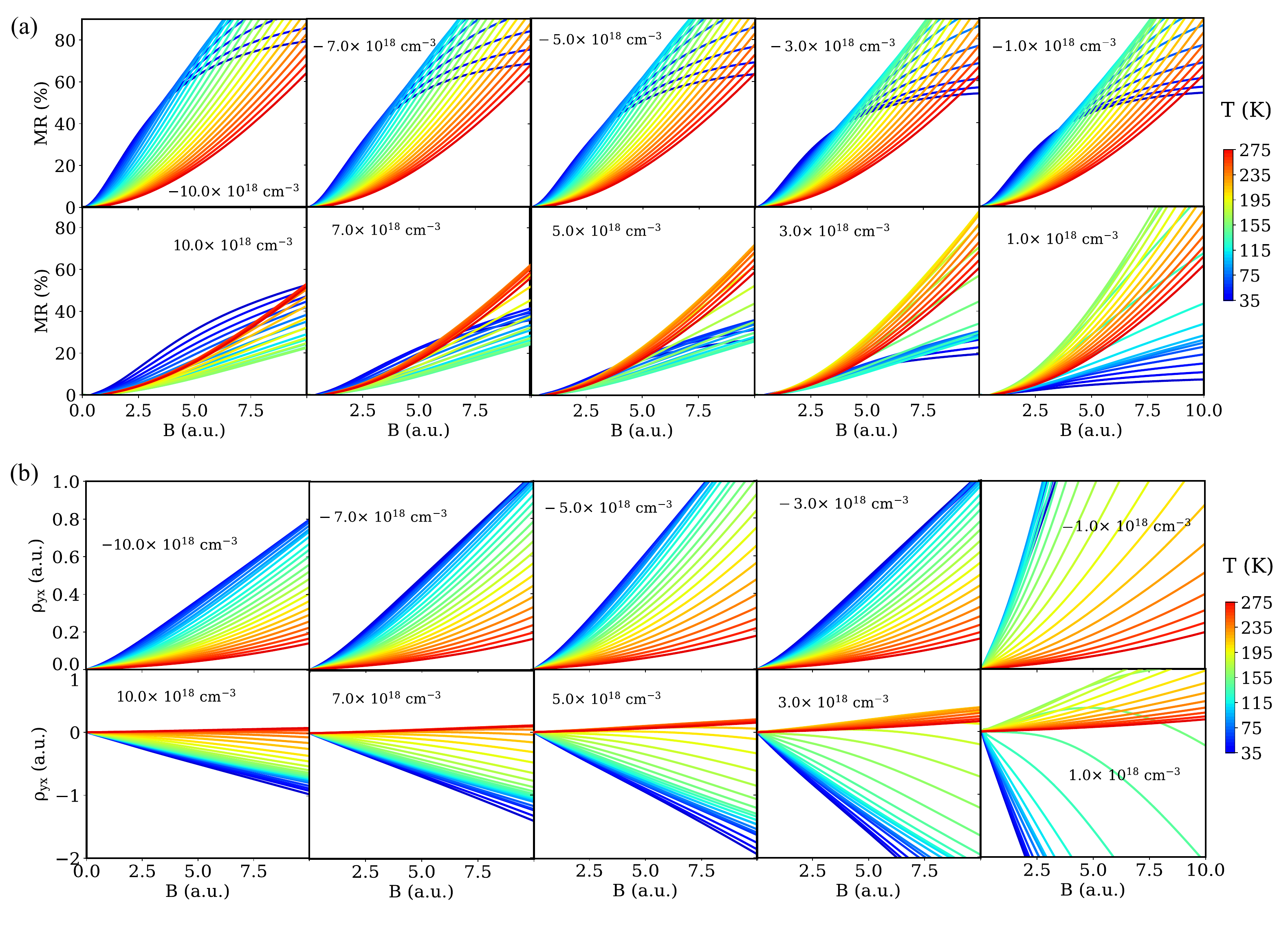}
    \caption{  The field-dependent (a) MR and (b) Hall resistivity at different temperatures with the fixed charge carrier concentration ranging from $n = -10.0\times 10^{18}\ \text{cm}^{-3}$ to $10.0\times 10^{18}\ \text{cm}^{-3}$. Temperature is varied from 35 K to 275 K at steps of 10 K. The relation between the temperature and color is indicated in the color bar.\label{figs2}}
\end{figure*}

\nocite{*}

\bibliography{refs}
\end{document}